\documentclass[twoside]{article}
\usepackage[accepted]{aistats2016}
\usepackage{amsmath,amssymb,fullpage,graphicx,mathptmx}
\let\hat\widehat
\let\tilde\widetilde

\newtheorem{theorem}{Theorem}
\newtheorem{lemma}[theorem]{Lemma}

\newenvironment{proof}{{\bf Proof.}}{$\Box$}

\renewcommand{\P}{\mbox{$\mathbb{P}$}}
\newcommand{\E}{\mbox{$\mathbb{E}$}}
\newcommand{\R}{\mbox{$\mathbb{R}$}}

\newcommand{\cD}{\mathcal{D}}

\begin{document}

%

%

\twocolumn[

\aistatstitle{Statistical Analysis of Persistence Intensity Functions}

\aistatsauthor{ Yen-Chi Chen \And Daren Wang \And Alessandro Rinaldo \And Larry Wasserman}

\aistatsaddress{ Carnegie Mellon University\\ Department of Statistics } 

]

\begin{abstract}
Persistence diagrams are 
two-dimensional plots that summarize
the topological features of functions and are an important part
of topological data analysis.
A problem that has received much attention is how
deal with sets of 
persistence diagrams.
How do we summarize them, average them or cluster them?
One approach --- the persistence intensity function ---
was introduced informally by Edelsbrunner, Ivanov, and Karasev (2012).
Here we provide a modification and formalization of this approach.
Using the persistence intensity function, we can visualize
multiple diagrams, perform clustering
and conduct two-sample tests.
\end{abstract}

\section{Introduction}

Topological data analysis (TDA) is an emerging area in
statistics and machine learning
\cite{carlsson2009topology,fasy2014confidence}.
The goal of TDA is to extract useful topological information 
from a given data and make inferences using these information.
TDA has been applied various fields such as pattern recognition
\cite{chepushtanova2015persistence},
computer vision \cite{ghrist2008barcodes}, and biology
\cite{nicolau2011topology, nanda2014simplicial}.

One of the most popular and powerful approaches to TDA 
is persistent homology.
This is a multiscale approach that finds topological features at various scales.
The result is summarized in a
two dimensional plot called a \emph{persistence diagram}
\cite{cohen2007stability, edelsbrunner2008persistent,fasy2014confidence,
edelsbrunner2002topological, zomorodian2005computing}.
Given a data set 
${\cal X} = \{X_1,\ldots, X_n\}$ with $X_i\in\mathbb{R}^d$,
we first find a summary function $f$.
Examples include the distance function
$f(x) = \min_i ||x-X_i||$
or the kernel density estimator
$f(x) = (n h^d)^{-1}\sum_i K( (X_i-x)/h)$
where $K$ is a kernel and $h$ is a bandwidth.
The persistence diagram summarizes the topological information of the level sets
of $f$ and converts it into points that can be represented by a 2-D diagram ${\cal D}$.
For distance functions, one uses the lower level sets.
For density estimator, one uses the upper level sets.
The function $f$ is a random function.
Hence, the corresponding persistence diagram is also random.

Now suppose we have $N$ datasets ${\cal X}_1,\ldots, {\cal X}_N$
giving rise to $N$ summary functions
$f_1,\ldots, f_N$
which in turn give rise to
diagrams
${\cal D}_1,\ldots, {\cal D}_N$.
The following question now arises: how do we summarize the diagrams?
For example how do we average the diagrams or cluster the diagrams?
One approach is based on the notion of Frechet means
\cite{mileyko2011probability, turner2014frechet}.
Here, the space of all persistence diagrams is endowed with a metric
(called the bottleneck distance).
The space is very complicated and the Frechet mean is a way to define
averages in this space.
This method is mathematically rigorous but is very complex is not easy for the
user to understand.

The second method is based on converting the
diagram into a set of one dimensional functions
called landscapes \cite{bubenik2012statistical}.
The statistical properties of landscapes
were further investigated in
\cite{chazal2013bootstrap,chazal2014stochastic,fasy2014confidence}.

The landscape approach is quite simple but
the results can be hard to interpret.
Perhaps the most appealing method is
due to 
\cite{edelsbrunner2012current}.
They introduce the intensity function
which is very easy to interpret.
They divide the plane into cells, the count the number of points
in each cell.
This converts the persistence diagram into a two-dimensional histogram.
The resulting function is easy to visualize, and can be averaged and clustered
very easily.
Similar ideas were introduced by
\cite{reininghaus2014stable, chepushtanova2015persistence,Pranav_thesis}.

The purpose of this paper is to make the intensity function
approach more rigorous.
We also modify the approach: instead of using histograms (cell counts)
we use kernel smoothing which leads to much smoother summaries.

The $N$ diagrams now give rise to $N$ two-dimensional
intensity functions
$\kappa_1,\ldots, \kappa_N$.
Mathematically, we regard these as $N$ random functions
drawn from some measure $\Omega$.
The mean
$\mathbb{E}_\Omega (\kappa_i)$
gives a well-defined population quantity.
We then show that
we can visualize the relative proximity of diagrams
and perform clustering for diagrams and
conduct a two sample test for two sets of diagrams.

\begin{figure*}
\center
\includegraphics[width=2in]{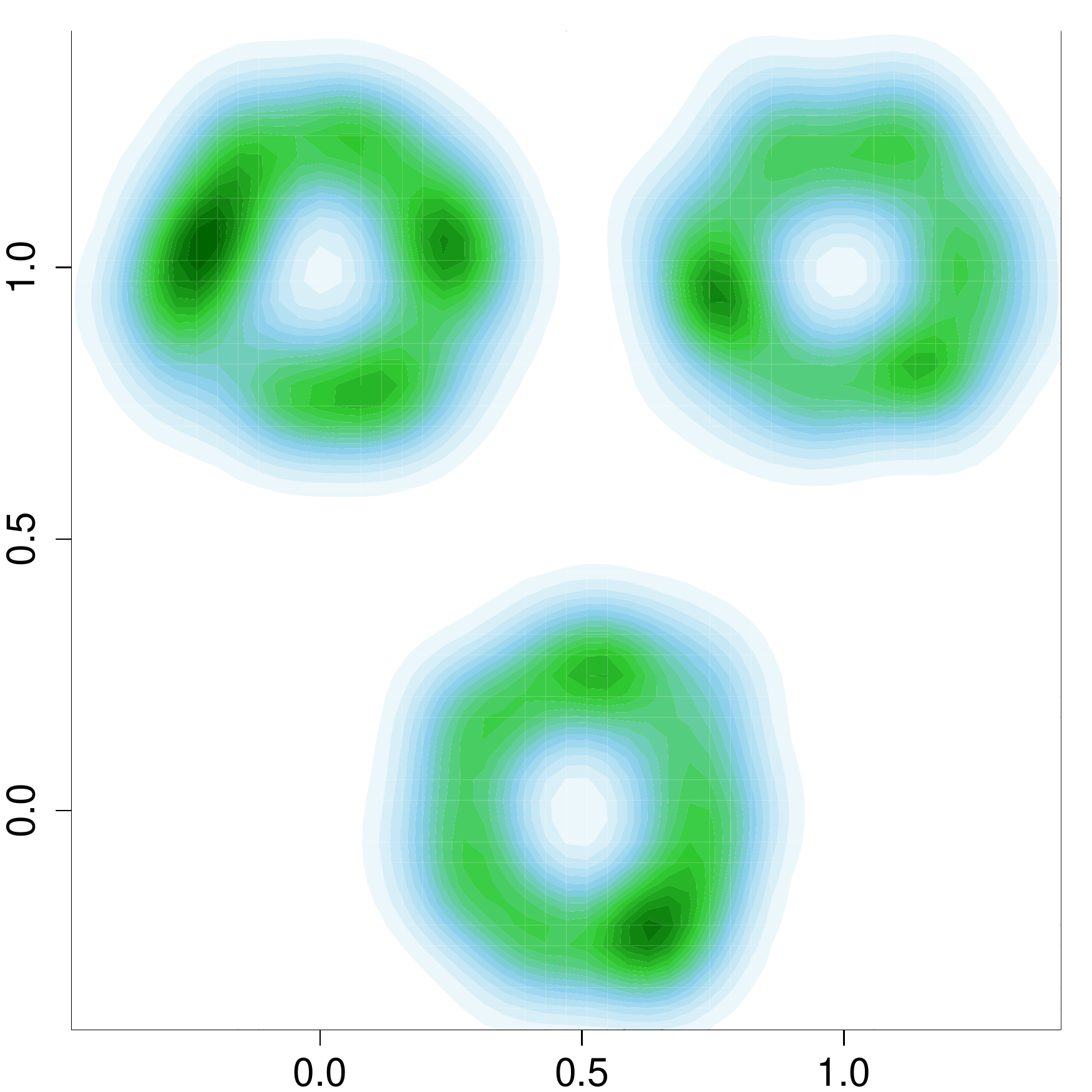}
\includegraphics[width=2in]{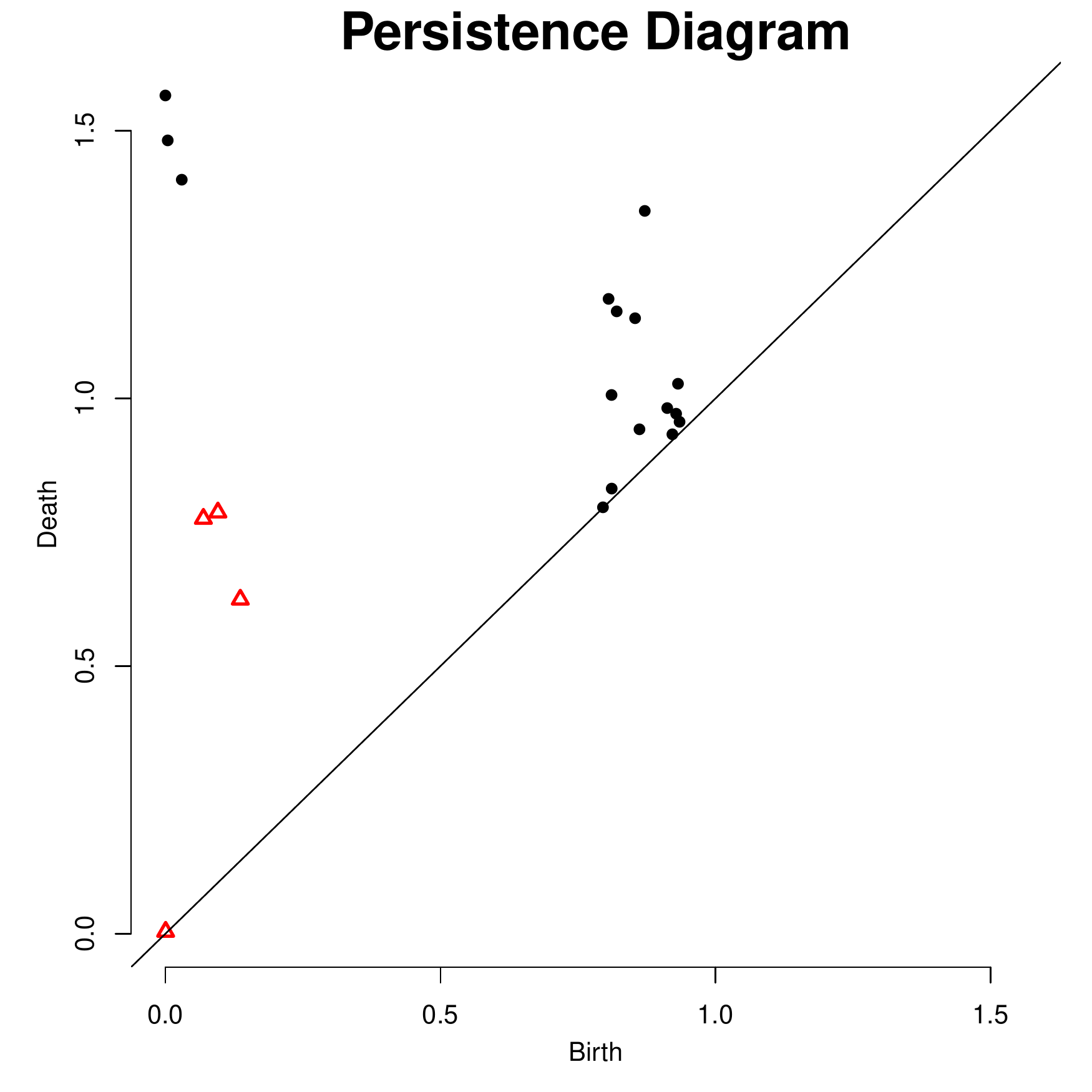}
\includegraphics[width=2in]{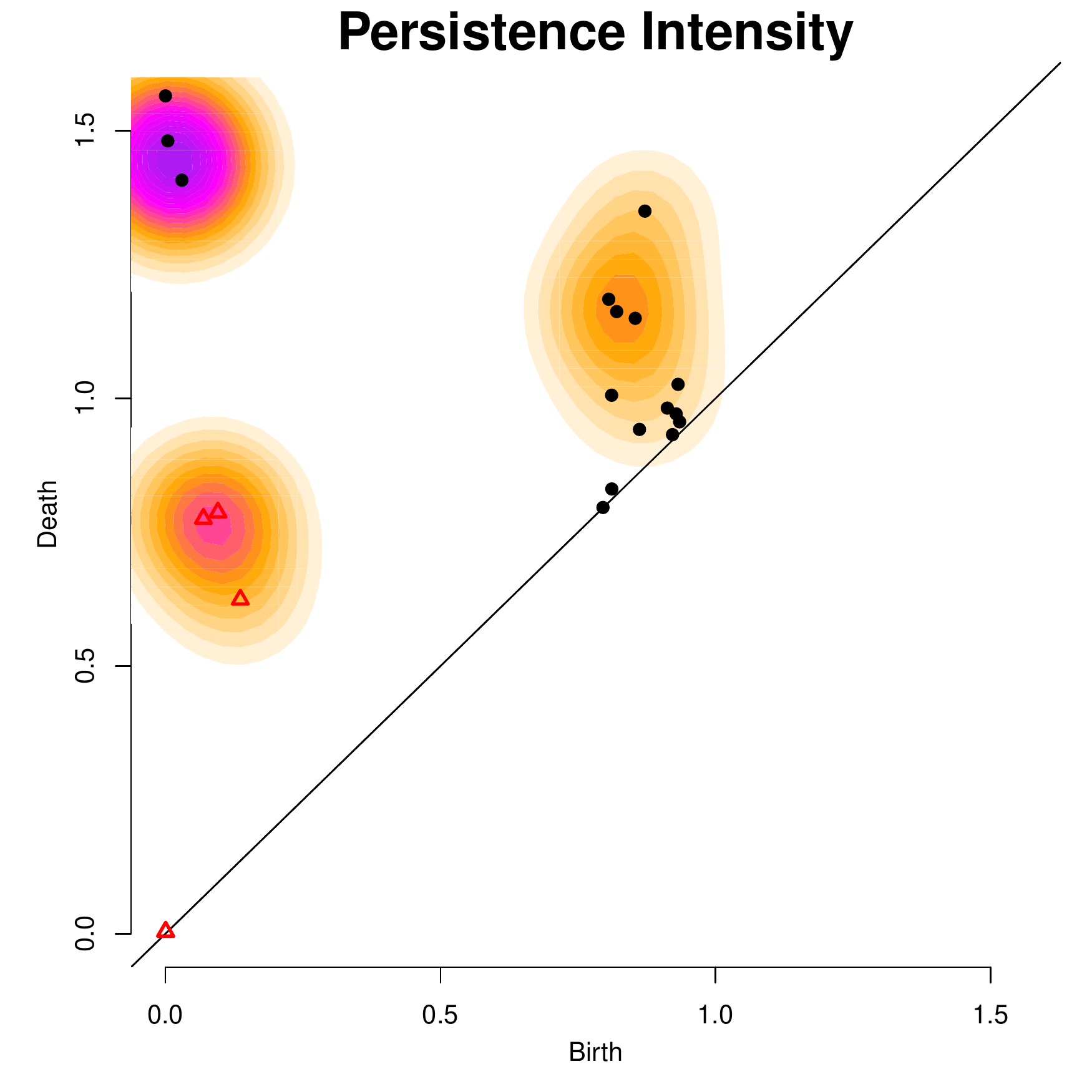}
\caption{An example of a persistence diagram and 
the smoothed persistence intensity estimator
constructed from a density estimator.
Left: the density estimator.
Middle: the persistence diagram.
Each black dot is a $0$-dimensional topological feature
and each red triangle is a $1$-dimensional topological feature.
Right: the smoothed persistence intensity estimator.
Note that in this case we only use the topological feature of dimension $0$ to
compute the intensities. i.e. 
the connected components.}
\label{fig::ex1}
\end{figure*}

\section{The Persistence Intensity Function}

We begin with a short introduction to the persistent homology
\cite{edelsbrunner2002topological, zomorodian2005computing}.
Given a function $f(x)$, 
let the upper level set be $L_t = \{x: f(x)\geq t\}$.
$L_t$ changes as the level $t$ varies.
Persistent homology studies how the topological features
of $L_t$ change when the level $t$ changes.
Some common topological features include
connected components (the zeroth order homology),
loops (the first order homology), voids (the second order homology), etc.
For more detailed discussion and definitions for the persistence homology,
we refer to \cite{edelsbrunner2008persistent}.

When the level $t$ varies, new topological features
may be created (birth) and 
some existing features may be eliminated (death).
The persistence diagram is a diagram that keep track of the birth and the death
of each topological features.
See the left and middle panel of Figure~\ref{fig::ex1} for an example
of persistence diagram generated from a function.
Formally, the \emph{pth persistence diagram}
$\cD(f)$ is the collection of points in the extended plane $\bar{\R}^2$,
where $\bar{R} = R\cup\{-\infty, +\infty\}$ such that
each point $(x,y)$ in the diagram 
represents a distinct $p$-th order topological feature that are created
when $t=x$ and is destroyed when $t= y$.
Thus, each persistence diagram can be presented as
a collection of points $\{(b_j, d_j): j=1,\cdots, K\}$ such that
each $(b_j,d_j)$ denotes the birth time and death time of $j$-th feature.

{\bf Remark:}
One usually defines persistent homology in terms of lower level sets.
But for probability density functions,
upper level sets are of more interest.
The result is that the points will appear below the diagonal
on the diagram.
However, is it customary to put the points above the diagonal.
Thus we simply switch the birth and death times which gives the usual diagram
although, techincally, we have births coming after deaths.
Moreover, we ignore the zeroth order topological feature with longest
life time $d_j-b_j$ since this feature represents the entire space;
it has infinite lifetime and contains no useful information.

Given a (random) function $f$ generated from a dataset ${\cal X}$ 
that is sampled from a distribution $P$,
its persistence diagram 
$$
\cD \equiv \cD(f)= \{(b_j,d_j): j=1,\cdots, K\}
$$
is a random object.
The random function $f$ is typically generated from 
density estimates,
or the distance function \cite{fasy2014confidence}
or something called the distance to a measure 
\cite{chazal2014robust, chazal2011geometric}
of ${\cal X}$ .

Our first step is to convert the persistence diagram
into a measure.
We define a random measure using $\cD$ as 
$\Phi(x,y) = \sum_{j=1}^K w_j\delta_{b_j, d_j}(x,y)$, 
where
$\delta_{a,b}(x,y)$ is a point mass at $(x,y) = (a,b)$
and $w_j$ is a weight.
In this paper, we take the weight to be
$w_j = d_j-b_j$. This places less weight to features near the diagonal which are features
with very short lifetimes.
Other weights can be used as well.
Now, $\Phi(x,y)$ is a random measure in $\R^2$ such
that the point mass located at $(b_j,d_j)$
has weight $(d_j-b_j)$ which equals its lifetime.

For each Borel set $B$ of $\R^2$, the persistence intensity measure is defined as
$$
R_P(B) = \E_P(\Phi(B)) = \E_P\left(\int_B \Phi(x,y)dxdy\right)
$$
where the expectation is with respect to $P$.
The \emph{persistence intensity function}  (or persistence intensity)
\begin{equation}
\kappa_P(x,y) = \lim_{\tau\rightarrow0} \frac{R_P(B((x,y),\tau))}{\pi\tau^2},
\label{eq::int}
\end{equation}
where $B((x,y),\tau)$ is a ball center at $(x,y)$ with radius $\tau$.
The persistence intensity at position $(x,y)$ 
is a weighted intensity for the presence of a topological at the given location.
The persistence intensity function is the parameter of interest to us.
Note that the persistence intensity is zero below the diagonal,
so smoothing estimators would suffer from boundary bias (see, e.g., page 73
of \cite{wasserman2006all}) along
the diagonal.
By adding the weight that is proportional to the life time,
we can alleviate this issue since points around diagonal (boundary)
are given a very low weight.

However, in practice we do not know $P$ so that 
$\kappa_P(x,y)$ has to be estimated.
Given a diagram $\cD = \{(b_j, d_j)\}_{j=1}^K$,
we estimate the intensity by
$$
\hat{\kappa}_\tau(x,y) = \sum_{j}(d_j-b_j)\frac{1}{\tau^2} K\left(\frac{x-b_j}{\tau}\right)K\left(\frac{y-d_j}{\tau}\right),
$$
where the sum is over all features,
$K$ is a symmetric kernel function such as a Gaussian and $\tau$
is the smoothing parameter.
In other words, $\hat{\kappa}_\tau(x,y)$ is just the smoothed version of
the persistence measure $\Phi(x,y)$.
Figure~\ref{fig::ex1} presents an example for a single diagram.

\begin{figure*}
\center
\includegraphics[height=1in]{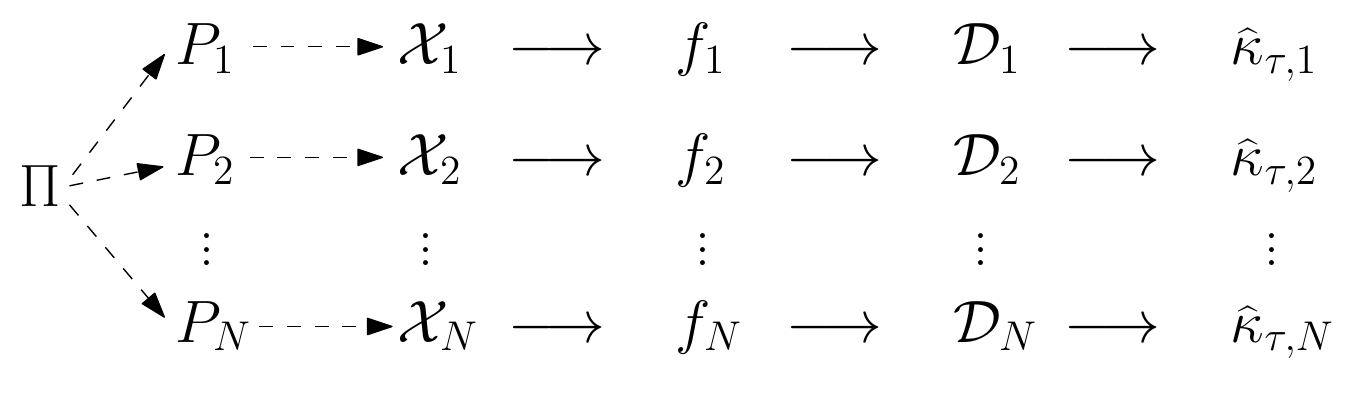}
\caption{A diagram showing how each quantity is connected to each other.
The dashed arrows from $\Pi$ to $P_j$ and from $P_j$ to ${\cal X}_j$
indicate the random sampling processes.
The solid arrows `$\longrightarrow$' indicate a (deterministic) derivation
from one quantity to another.
}
\label{fig::diagram}
\end{figure*}

Now we consider the case that we have multiple datasets 
${\cal X}_1, \cdots, {\cal X}_N$.
The data in data set ${\cal X}_j$
come from a distribution $P_j$.
The distributions $P_1,\ldots, P_N$
are themselves random draws from some distribution
which we denote by $\Pi$.
Namely, we 
use $\Pi$ to generate $P_j$,
and sample from $P_j$ to obtain ${\cal X}_j$,
and finally use ${\cal X}_j$ to construct $f_j$.
In this case, $f_1,\cdots,f_N$ are iid random functions from some measure $\Omega$.
Figure~\ref{fig::diagram} shows a diagram for how these quantities are 
generated from one another.
Note that there are two sources of randomness: $\Pi$ and each $P_j$.
It is easy to see that if $P_j$ is given, the randomness of $f_j$,
the persistence diagram $\cD_j$,
and its associated random measure $\Phi(x,y)$ 
are determined by $P_j$. 
For each distribution $P_j$, we have a persistence intensity by
equation \eqref{eq::int}
$$
\kappa_{j}(x,y)\equiv \kappa_{P_j}(x,y)
$$
and using the given diagram $\cD_j$, we have an estimator
$$
\hat{\kappa}_{\tau,i}(x,y).
$$
Note that $\kappa_j$ depends on $P_j$ so it is actually a random quantity.
However, we cannot consistently estimate each individual $\kappa_j$
since we only have one diagram.
On the other hand, if we consider the population version of intensity function,
we can consistently estimate it.

There are two ways to define a population intensity function
and later we will show that they are the equivalent.
For the first definition, we define
$$
\bar{\kappa}(x,y) = \E_{\Pi}(\kappa_j(x,y)).
$$
Namely, $\bar{\kappa}(x,y)$ is the average intensity 
using the distribution $\Pi$.
Alternatively, we consider the distribution $\tilde{P} = \E_{\Pi}(P_j)$,
which is the `mean' distribution of $\Pi$.
Let $\tilde{{\cal X}}$ be a sample from $\tilde{P}$.
Then we define
$$
\tilde{\kappa}(x,y) \equiv \kappa_{\tilde{P}}(x,y)
$$
by equation \eqref{eq::int}.
Both $\bar{\kappa}(x,y)$ and $\tilde{\kappa}(x,y)$
are population level quantity.
The following lemma shows that they are the same.
\begin{lemma}
Let $\tilde{\kappa}(x,y), \bar{\kappa}(x,y)$ be defined as the above.
Assume $\tilde{\kappa}(x,y), \bar{\kappa}(x,y)$ are bounded.
Then
$$
\tilde{\kappa}(x,y)=\bar{\kappa}(x,y).
$$
\label{lem::equiv}
\end{lemma}
We always assume both $\tilde{\kappa}(x,y),\bar{\kappa}(x,y)$
are bounded
so they are the same.
For simplicity we write
\begin{equation}
\kappa(x,y) \equiv \tilde{\kappa}(x,y)=\bar{\kappa}(x,y).
\end{equation}
A nonparametric estimator for the population intensity function $\kappa(x,y)$ is
\begin{equation}
\hat{\kappa}_n(x,y) = \frac{1}{n}\sum_{i=1}^n \hat{\kappa}_{\tau,i}(x,y).
\end{equation}
Essentially, this is the sample average for the intensity function.

One may consider other weights
such as
$$
w_j = g(\alpha_j)L_{\alpha_j}(d_j-b_j),
$$
where $\alpha_j$ is the dimension of $j$-th topological feature
and $L_k$ is some smooth function with $L_k(0)=0$.
The reason we impose the constraint $L_k(0)=0$ is to 
avoid a discontinuity of $\kappa(x,y)$ along the diagonal
since we will not have any topological features below the diagonal line.
The function $g$ determines how we want to give different weights
to topological features with different dimensions
and each $L_k$ is a function that determines how we want to give
weight to the $k$ dimensional topological features according to their lifetime.
Note that the parameter of interest $\kappa(x,y)$ 
depends on the weight we choose.
In this paper, we use life time as the weight $w_j=d_j-b_j$, which is 
the case that $g(\alpha_j) = 1$ and $L_k(x) = x$.
Although our theoretical analysis is done for the this simple case,
it can be generalized to other weights easily.

\section{Statistical Analysis}

Here we study the asymptotic behavior of $\hat{\kappa}_N(x,y)$.
Throughout this paper, we assume all random functions
are Morse functions 
\cite{milnor1963morse, morse1925relations,morse1930foundations}.
This means that the Hessian is non-degenerate at its critical points
\cite{cohen2007stability, chazal2014stochastic,chazal2014robust}.

For a univariate function $g$, let $g^{(s)}$ denote its $s$-th derivative.
We make the following assumptions.

\begin{enumerate}
\item[(A1)] The persistence intensity function $\kappa(x,y)<\infty$
and has at least two bounded continuous derivatives.

\item[(A2)] $\int\kappa(x,y)dxdy<\infty$ and $\int \nabla^2\kappa(x,y) dxdy <\infty$.

\item[(K)] The kernel function $K$ is symmetric, at least bounded
  twice continuously differentiable. Moreover, $\int K(x)dx = 1, \int
  K^{(2)}(x)dx <\infty$.
\end{enumerate}

(A1) requires the persistence intensity
to be well defined and smooth.
Assumption (A2) is to regularize the integrated behavior
for $\kappa(x,y)$ and its derivatives. 
Assumption (K) is a common assumption for kernel smoothing method
such as the kernel density estimation and kernel regression 
\cite{wasserman2006all,tsybakov2008introduction}.

A direct result from assumption (A1) is the following useful lemma.
\begin{lemma}
Assume (A1) for $\kappa_P$ and consider
the case where the sample ${\cal X}$ is from $P$. Then
for any twice differentiable function $h(x,y)$,
\begin{equation}
\begin{aligned}
\E\int h(x,y)&\Phi(x,y)dxdy \\
& \equiv \mathbb{E}\left(\sum_{j=1}^K (d_j-b_j)h(b_j,d_j)\right)\\
&= \int h(x,y) \kappa_P(x,y)dxdy.
\end{aligned}
\end{equation}
\label{lem::bias1}
\end{lemma}
Lemma \ref{lem::bias1} shows that the expectation of integration
(first expression) is equivalent to integration of expectation (last expression).

The next result, which is based on Lemma~\ref{lem::bias1},
controls the bias for the smoothed persistence intensity estimator.
\begin{lemma}
Assume (A1) for $\kappa_P$ and assume (K)
and consider the case where the sample ${\cal X}$ is from $P$.
Then 
the bias for estimating $\kappa_P(x,y)$ using a single smoothed function $\hat{\kappa}_\tau(x,y)$ is
\begin{align*}
\E(\hat{\kappa}_\tau(x,y)) - \kappa_P(x,y)
= C_0\cdot\nabla^2\kappa_P(x,y)\cdot\tau^2 +  o(\tau^2),
\end{align*}
for some constant $C_0$ that depends only on the kernel function $K$.
\label{lem::bias2}
\end{lemma}
This bias is essentially the same as the bias for nonparametric density estimation;
see, e.g., page 133 of \cite{wasserman2006all}.
Lemma~\ref{lem::bias2} shows that the bias 
is small for a small $\tau$, 
and the variance of $\hat{\kappa}_\tau(x,y)$
is of the order of $O(\tau^{-2})$.

Now we prove that the sample average estimator $\hat{\kappa}_N(x,y)$ is a
consistent estimator of the population $\kappa(x,y)$
under the mean integrated square error (MISE) 
\cite{tsybakov2008introduction}.
\begin{theorem}
Assume (A1--2) and (K).
Then mean integrated square error for $\hat{\kappa}_N(x,y)$
\begin{align*}
\E\int\left(\hat{\kappa}_N(x,y)- \kappa(x,y)\right)^2 dxdy
&= O(\tau^4) + O\left(\frac{1}{N\tau^2}\right),
\end{align*}
which implies the optimal smoothing bandwidth
$$
\tau^*_N = O(N^{-1/6}).
$$
\label{thm::MISE}
\end{theorem}
Theorem~\ref{thm::MISE} gives
the mean integrated square error rate for 
the estimator $\hat{\kappa}_N(x,y)$
and also the optimal rate of smoothing parameter $\tau$
when we have more and more diagrams.
An interesting feature for Theorem~\ref{thm::MISE}
is that this rate is the same as the kernel density estimator
in dimension $2$ \cite{wasserman2006all}.
This is reasonable since the smoothed intensity function
is to smooth a measure in 2-D.
Note that assumptions (A1) and (K) together are enough for
the consistency for the (pointwise) mean square error 
$\E\left(\hat{\kappa}_N(x,y)- \kappa(x,y)\right)^2$
and the rate is at the same order of MISE.

Before we end this section, we show the
asymptotic normality of $\hat{\kappa}_N(x,y)$.
\begin{theorem}
Assume (A1--2) and (K).
Then 
$$
\sqrt{N\tau^2}\left(\hat{\kappa}_N(x,y)-\E(\hat{\kappa}_N(x,y))\right)
\overset{d}{\rightarrow} N(0, \sigma^2(x,y)),
$$
where $\overset{d}{\rightarrow}$ denotes the convergence in distribution
and $\sigma^2(x,y)$ is some function depending only on $\kappa(x,y)$ 
and the kernel function $K$.
\label{thm::normal}
\end{theorem}
The proof to this theorem is simply an application of the central limit theorem
so we omit it.

\section{Applications of Intensity Functions}

The power of the persistence intensity function is that
the smoothing technique converts
a persistence diagram into a smooth function
that carries topological information of the original function.
Thus, we can apply methods for comparing functions
to analyze the similarity between diagrams.

\subsection{Clustering and Visualization}	\label{sec::vis}

\begin{figure}
\center
\includegraphics[width=3in]{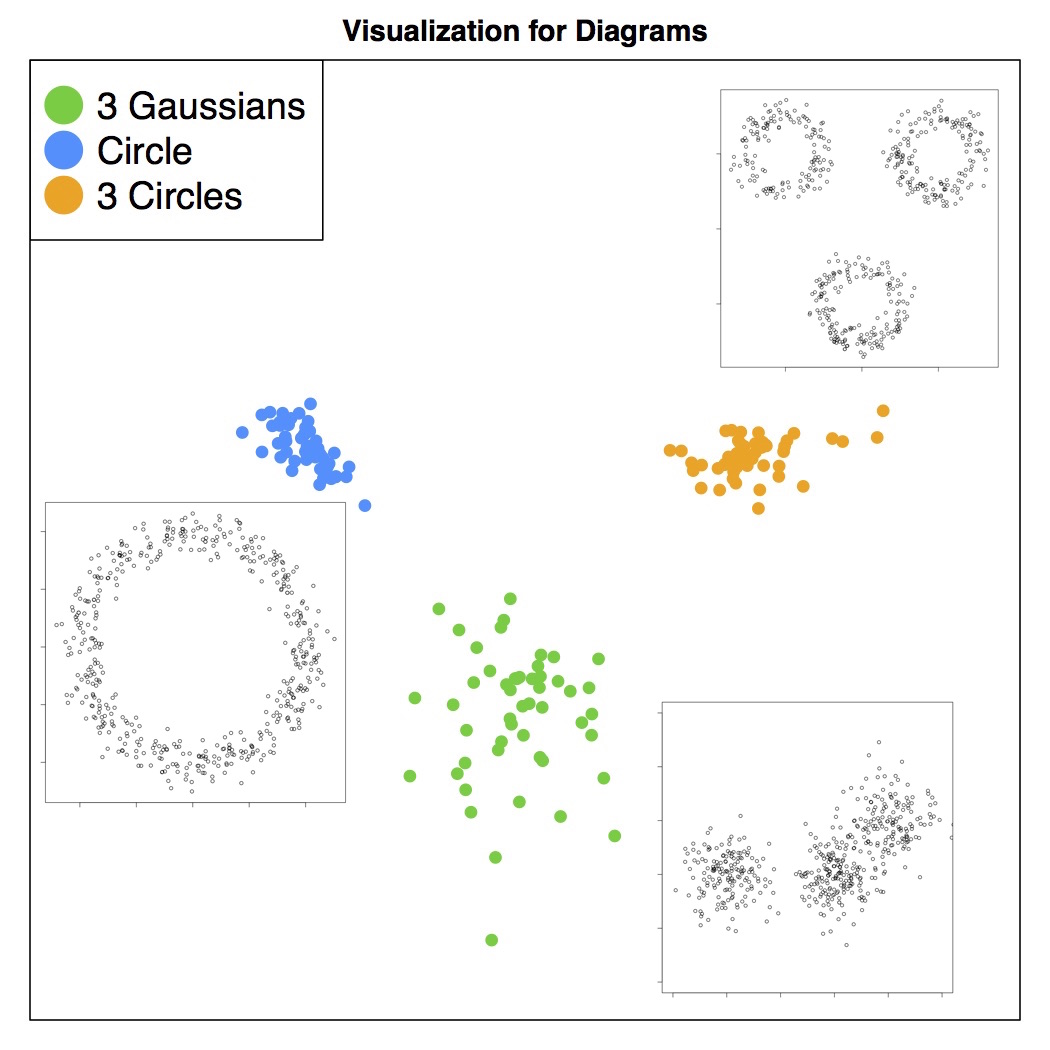}
\caption{Visualization for multiple diagrams using persistence intensity function.
We generate $50$ diagrams per each of the three types of data: one circle (blue), 
three circle (orange), and three Gaussian mixture (green).
We see a clear separation between each of the three types of point clouds.
The detailed implementation is given in Section~\ref{sec::vis}.
}
\label{fig::vis}
\end{figure}

\cite{chazal2009gromov} used persistent homology
to cluster images.
They computed the bottleneck distance 
between each pair of diagrams and mapped the points to the plane
using 
multidimensional scaling (MDS) \cite{kruskal1964multidimensional}
to visualize the results.
Here we consider the same idea using 
persistence intensity functions.

For pairs of diagrams, say $\cD_1$ and $\cD_2$,
we first compute
their smoothed persistence intensity $\hat{\kappa}_{\tau,1}(x,y)$
and $\hat{\kappa}_{\tau,2}(x,y)$
and then calculate their pairwise difference $\Delta_{12}$ using
some loss functions such as $L_1$ loss.
Namely,
$$
\Delta_{12} = \int |\eta_{\tau,1}(x,y) -\eta_{\tau,2}(x,y)|dxdy.
$$
This allows us to construct
a pair distance matrix $\Delta$ for diagrams $\cD_1,\cdots, \cD_N$.
Each element $\Delta_{ij}$
denotes the distance between diagram $\cD_i$ and diagram $\cD_j$.
Given $\Delta$,
we can use techniques from MDS
to visualize several diagrams.
This visualization shows the relative proximity among 
diagrams in terms of the persistence intensity.

Figure~\ref{fig::vis} presents an example for using persistence 
intensity to visualize diagrams generated from three different populations.
The three populations are the kernel density estimates
of three types of point clouds:  
one circle, three circles, and three Gaussian mixtures.
One circle is a uniform distribution with radius $r=1$ and corrupted with
a Gaussian noise with standard deviation $0.1$.
Three circles is a uniform random distribution around
three circles centered at $(0,0)$, $(1,0)$, and $(1.5,0.5)$
with radius $0.25$ and then we add Gaussian noise with standard
deviation $0.05$.
The three Gaussian mixture is three Gaussians
centered at $(0,0)$, $(1,0)$, and $(1.5,0.5)$ with standard deviation
$0.2$ along both axes and zero correlation.
We generate the point cloud with size $500$ points
and then use kernel density estimator with Gaussian kernel and 
smoothing parameter $h=0.07$.
Thus, each random function is an estimated density function
for a point cloud.
For each of the three types of data, we generate 50
point clouds, so we have $150$ random functions (density estimates).
We then compute the persistence diagrams for
the $150$ random functions 
and smooth the diagrams, using $\tau=0.1$, into intensity functions.
Given the intensity functions, we use $L_1$ distance to construct
the pairwise distance matrix $\Delta$ and then use classical multidimensional scaling (MDS)
\cite{kruskal1964multidimensional}
to visualize the data.

\begin{figure}
\center
\includegraphics[width=3in]{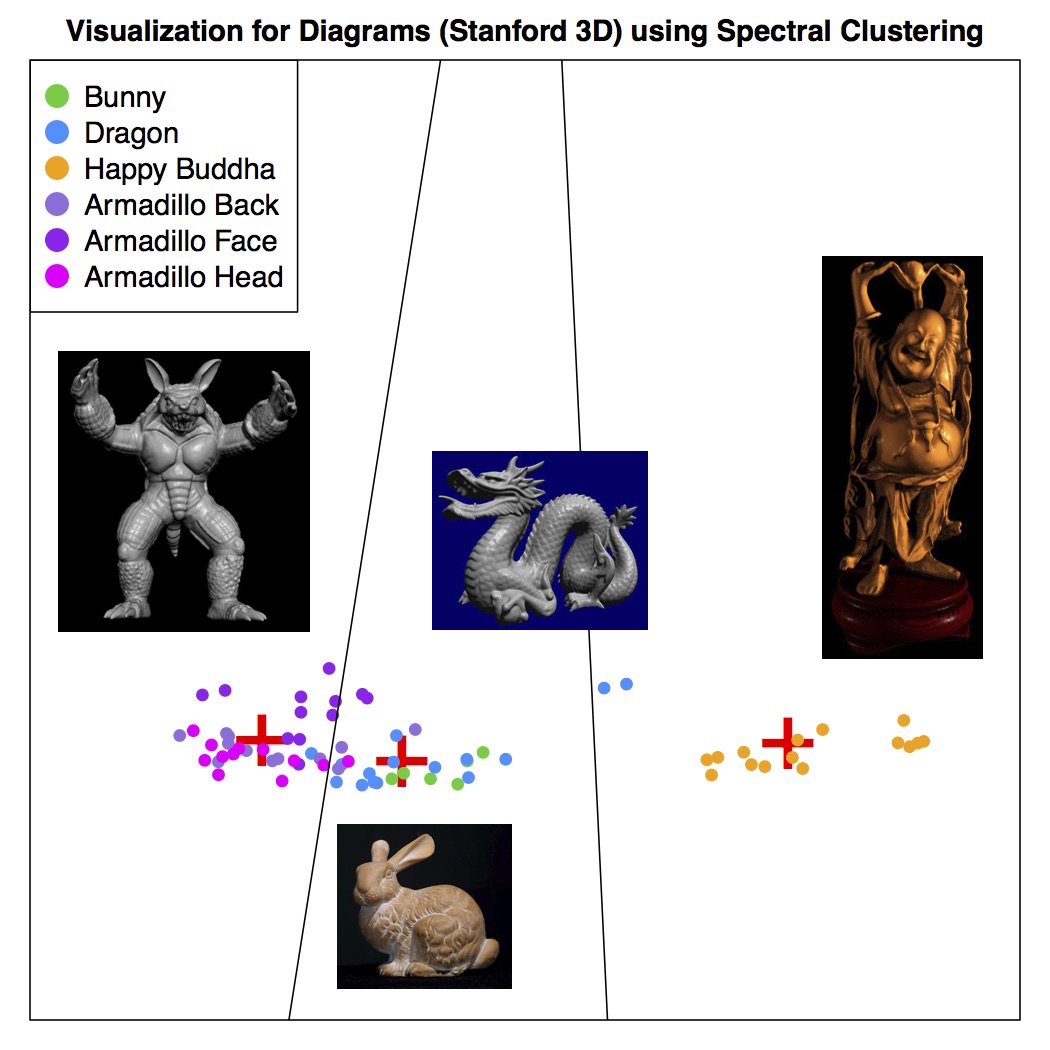}
\centering
\begin{tabular}{rrrr}
  \hline
 & 1 & 2 & 3 \\ 
  \hline
Armadillo Back &   9 &   3 &   0 \\ 
  Armadillo Face &  10 &   2 &   0 \\ 
  Armadillo Head &  11 &   1 &   0 \\ 
  Bunny &   0 &   6 &   0 \\ 
  Dragon &   1 &  12 &   2 \\ 
  Happy Buddha &   0 &   0 &  15 \\ 
   \hline
\end{tabular}
\caption{Visualization and clustering for data from the 
Stanford 3D Scanning Repository.
The top figure visualize
persistence diagrams for different point clouds using
the two smallest eigenvectors
from spectral clustering.
The three red crosses denote the cluster center from k-means clustering 
using $k=3$
and the two straight lines denote the cluster boundaries.
The bottom table shows the confusion matrix for the clustering result.
The detailed implementation is given in the last
paragraph of Section~\ref{sec::vis}.
}
\label{fig::vis2}
\end{figure}

Moreover, 
we can use the idea of 
spectral clustering \cite{von2007tutorial}
to cluster and visualize these diagrams simultaneously.
Figure~\ref{fig::vis2} presents an example for 
the dataset from the Stanford 3D Scanning Repository \cite{curless1996volumetric}.
The dataset is a point cloud characterizing the 3D surfaces
of objects at different angles when taking 3D scanning.
Thus, each object has several point clouds
and each point cloud represent the 3D image of this object
at one particular angle.
We use the bunny, the dragon, the happy Buddha, and three different 
focuses of the Armadillo. 
For each point cloud, we use distance function and compute its persistence diagrams.
Then we smooth the persistence diagrams using $\tau=0.01$
and impose $5$ times weight for the first order homology features (loops)
to get the smoothed persistence estimator. 
We use $L_1$ distance to compare pairwise distance
and then convert the distance matrix $\Delta$ into a similarity matrix $S$ by 
$S_{ij} = e^{-\Delta_{ij}/200}$.
Finally we perform normalize cut spectral clustering over the
similarity matrix $S$.
We use the two eigenvectors (correspond to the two smallest eigenvalues) 
to visualize and cluster data points.
The visualization is given on the top panel of Figure~\ref{fig::vis2}.
It shows a clear pattern that
Armadillos (purple) aggregate at the left part;
the bunny (green) and the dragon (blue) are at the center of the figure;
the happy Buddha (orange) is on the right side.
We then perform k-means clustering \cite{hartigan1979algorithm} 
with $k=3$ to the first two eigenvectors.
The three red crosses denote the cluster center
and the two straight lines are the boundaries of clusters.
We give the confusion matrix at the bottom panel.
It shows that the first cluster contains majority of Armadillos;
the second cluster belongs to bunny and dragon;
the third cluster is dominated by Happy Buddha.

\subsection{Persistence Two Sample Tests}
Assume that we are given two sets of diagrams,
denoted as
$$
(\cD^{(1)}_1, \cdots, \cD^{(1)}_{n_1}),\,\,\,\, (\cD^{(2)}_1, \cdots, \cD^{(2)}_{n_2}).
$$ 
The goal is to determine if these two sets of diagrams
are generated from the same population or not.

A testing procedure based on the persistence intensity function is as follows.
Let
$\kappa_1(x,y)$
and $\kappa_2(x,y)$
be the persistence intensity for the two populations.
Then the null hypothesis is
$$
H_0: \kappa_1(x,y) = \kappa_2(x,y).
$$
Based on 
intensity estimators from both populations,
there are many ways to carry out the test.
Let $\hat{\kappa}_1(x,y)$ be the intensity estimator
based on data from first population and 
$\hat{\kappa}_2(x,y)$ be the intensity estimator
from second population.
One example for carrying the test is to use the $L_1$ distance
$$
T_{1} =\int \left|\hat{\kappa}_1(x,y) - \hat{\kappa}_2(x,y)\right| dxdy
$$
as a test statistics
and then perform permutation test 
(see, e.g., section 10.5 of \cite{wasserman2013all}) to get the p-value or 
bootstrap the test statistics to get the variance and convert it into z-score.

\begin{figure}
\center
\includegraphics[width=3in]{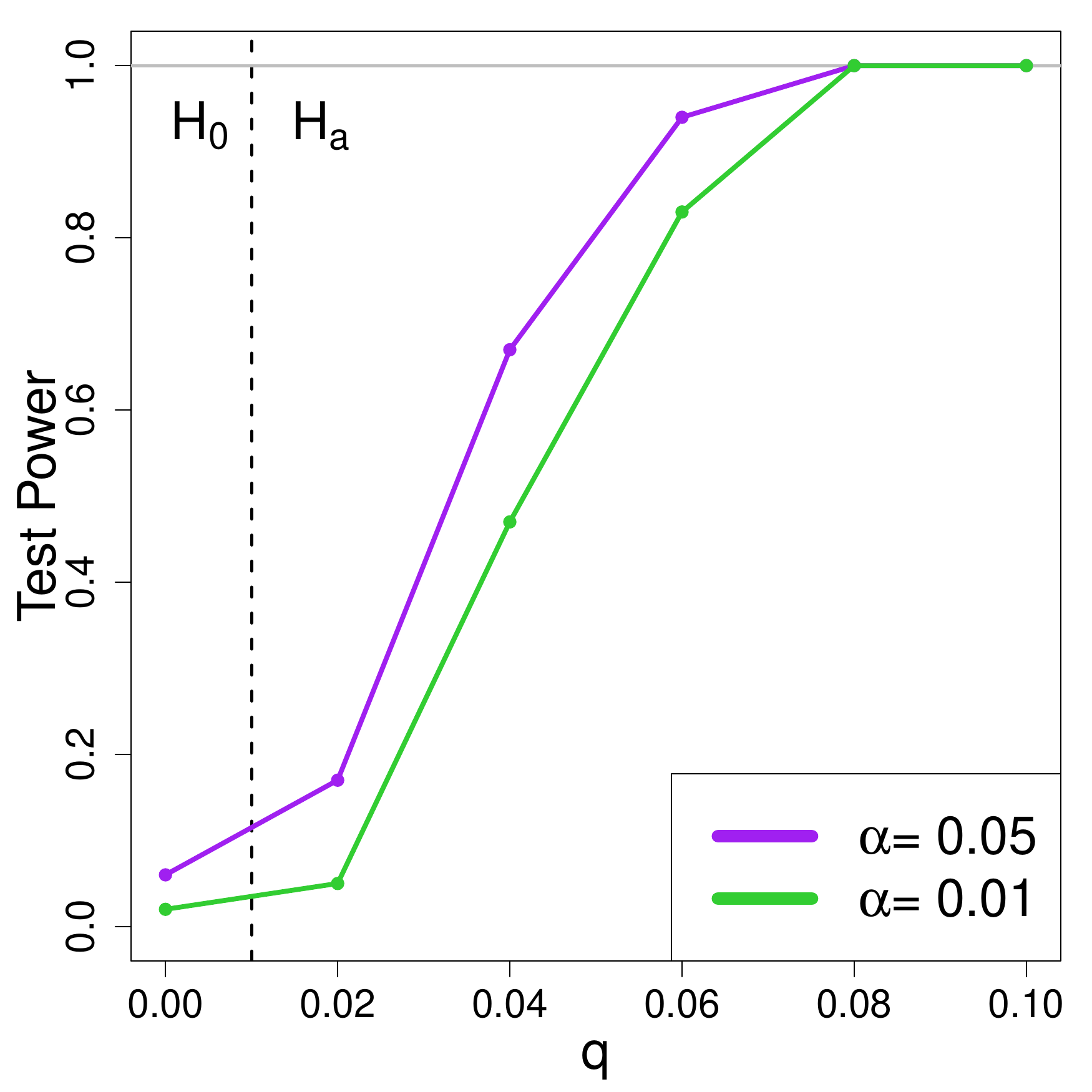}
\centering
\caption{
A simulation example for the two sample test.
The $x$ axis denotes difference between two distributions.
At $q=0$ both distributions are the same.
The purple and green colors denote the power for test under
significance $\alpha=0.05$ and $0.01$.
}
\label{fig::test}
\end{figure}

Figure~\ref{fig::test} presents an example for the two sample test.
We consider point clouds from two distributions:
\begin{align*}
P_1 &\sim {\sf Unif}[-1,1]^2,
\\P_2 &\sim (1-q)\cdot{\sf Unif}[-1,1]^2 + q \cdot {\sf P_C},
\end{align*}
where ${\sf P_C}$ is the uniform distribution over a circle with unit radius.
Namely, $P_1$ is just uniform distribution over the rectangle $[-1,1]^2$
and $P_2$ is the same as $P_1$ with probability $1-q$
and with probability $q$ it generates a random point on a unit circle.
We consider $q = 0, 0.02, 0.04,0.06,0.08,0.10$. Note that
when $q=0$, the two distributions are the same so that the 
power of the test should be the same as its significance level.
We generate $n=500$ points for each point cloud
and generate $N=50$ number of point clouds for each distribution.
Then we apply kernel density estimator for each point cloud
with smoothing parameter $h=0.1$ to obtain persistence diagrams.
Finally, we use $\tau=0.025$ to smooth the diagram into
persistence intensities and compute the average persistence intensity for both sample
and then apply $L_1$ distance with $1000$ times permutation test to compute the 
p-values.
We consider two significance levels $\alpha=0.05$ (purple) and $\alpha=0.01$ (green).
In Figure \ref{fig::test}, the test at both significance level controls
type-I error ($q=0$ case).
Moreover, the power for the test converges to $1$ rapidly 
even when two distributions
are just slightly different ($q<0.10$ indicates that only 
less than roughly $10\%$ of the 
data points in each point cloud are from different distributions).

\section{Conclusion}

In this paper, we study the persistence intensity as a summary
for topological features over distribution of random functions.
We propose a smoothed estimator for the intensity function and
derive statistical consistence for this smoothed estimator.
The main idea is to smooth persistence diagrams into
functions. This smoothing technique also allows us to
visualize persistence diagrams, perform clustering and 
conduct two-sample tests.
Our examples suggest that
the intensity function contains useful topological information.


\appendix
\section{Proofs}
\begin{proof}[Proof for Lemma \ref{lem::equiv}]
Since $\tilde{\kappa}(x,y),\bar{\kappa}(x,y)<\infty$,
by dominated convergence theorem
\begin{align*}
\tilde{\kappa}(x,y) 
& = 
\kappa_{\tilde{P}}(x,y)\\
&= 
\lim_{\tau\rightarrow 0}\frac{1}{\pi\tau^2}\E_{\Pi}(\E_{P_j}(\Phi(B)))\\
&= 
\E_{\Pi}\left(\lim_{\tau\rightarrow 0}\frac{1}{\pi\tau^2}\E_{P_j}(\Phi(B))\right)\\
&=
\E_{\Pi}( \kappa_j(x,y))\\
&= \bar{\kappa}(x,y).
\end{align*}

\end{proof}

\begin{proof}[Proof for Lemma~\ref{lem::bias1}]

Note that $\kappa(x,y)$ can be written as
\begin{align*}
\kappa_P&(x,y)\\ 
&= \lim_{\tau\rightarrow0} \frac{R_P(B((x,y),\tau))}{\pi\tau^2}\\
&=\lim_{\tau\rightarrow0}  
\E\left(\int\frac{1}{\pi\tau^2}I_\tau(x-v,y-u)
\Phi(u,v)dudv\right),
\end{align*}
where $I_\tau(x-v,y-u) = I\left(\frac{|x-u|}{\tau}\leq 1, \frac{|y-v|}{\tau}\leq 1\right) $
Thus, by moving $h(x,y)$ into the expectation,
\begin{align*}
\int &h(x,y) \kappa_P(x,y)dxdy \\
&= \int 
\lim_{\tau\rightarrow0}  
\E\Bigg(\int\frac{1}{\pi\tau^2}h(x,y) \\
&\qquad \qquad \qquad \times I_\tau(x-v,y-u)
\Phi(u,v)dudv\Bigg)dxdy
\end{align*}

Since $h(x,y)$ is bounded, by dominated convergence theorem and Fubini's theorem,
we can exchange the limit, expectation, and integrations,
\begin{align*}
\int &h(x,y) \kappa_P(x,y)dxdy \\
&= 
\lim_{\tau\rightarrow0}  \int 
\E\Bigg(\int\frac{1}{\pi\tau^2}h(x,y) \\
&\qquad \qquad \qquad \times I_\tau(x-v,y-u)
\Phi(u,v)dudv\Bigg)dxdy\\
& = 
\lim_{\tau\rightarrow0}  
\E\Bigg(\int \int\frac{1}{\pi\tau^2}h(x,y) \\
&\qquad \qquad \qquad \times I_\tau(x-v,y-u)
dxdy \Phi(u,v)dudv\Bigg)
\end{align*}

Note that we have
\begin{align*}
\frac{1}{\pi\tau^2}\int h(x,y)&I_\tau(x-v,y-u)dxdy\\
&= \int_{-1}^1\int_{-1}^1 h(v+a\tau, u+b\tau)dadb\\
& = h(u,v) + C_0 \nabla^2h(u,v)\tau^2 + O(\tau^4)
\end{align*}
by using the change of variable 
$(x-u)/\tau=a, (y-v)/\tau=b$.
Plugging this into the previous equality proves the result.


\end{proof}

\begin{proof}[Proof for Lemma~\ref{lem::bias2}]

By definition
\begin{align*}
\mathbb{E}(&\hat{\kappa}_\tau(x,y))\\
& = \E\left(\sum_{j=1}^K (d_j-b_j) \frac{1}{ \tau^2}K\left(\frac{x-b_j}{\tau}\right)
K\left(\frac{y-d_j}{\tau}\right)\right)\\
& = \E\left( \frac{1}{ \tau^2}K\left(\frac{x-u}{\tau}\right)
K\left(\frac{y-v}{\tau}\right)\Phi(u,v)dudv\right).
\end{align*}
Now apply Lemma~\ref{lem::bias1} with 
$$
h(u,v) =  \frac{1}{ \tau^2}K\left(\frac{x-u}{\tau}\right)
K\left(\frac{y-v}{\tau}\right),
$$
we obtain
\begin{align*}
\mathbb{E}(&\hat{\kappa}_\tau(x,y))\\ 
&= \int  \frac{1}{ \tau^2}K\left(\frac{x-u}{\tau}\right)
K\left(\frac{y-v}{\tau}\right) \kappa_P(u,v)dudv\\
& = \int K\left(a\right)
K\left(b\right) \kappa_P(x-\tau a, y-\tau b)dadb\\
& = \kappa_P(x,y) + C_1\cdot\nabla^2 \kappa_P(x,y) \cdot\tau^2  +o(\tau^2) ,
\end{align*}
where $C$ is some constant.
Note that we use Talyor expansion to the Hessian in the last equality
and the first derivatives integrated to $0$ since the kernel function $K$ 
is symmetric.

\end{proof}

\begin{proof}[Proof for Theorem~\ref{thm::MISE}]

The mean integrated errors can be factorized into the following two terms
\begin{align*}
\E\int(\hat{\kappa}_N(x,y)&-\kappa(x,y))^2 dxdy \\
&= \int B(x,y)^2dxdy +\int V_N(x,y)dxdy,
\end{align*}
where 
\begin{align*}
B(x,y) &= \mathbb{E}(\hat{\kappa}_N(x,y)) - \kappa(x,y)\\
V_N(x,y)& = {\sf Var}(\hat{\kappa}_N(x,y)).
\end{align*}

Since $\kappa(x,y) = \E_{\Pi}(\kappa_j(x,y)) $
and
$
\E(\hat{\kappa}_N(x,y)) =\E_{\Pi}(\hat{\kappa}_j(x,y)),
$
the bias 
$$
B(x,y) = C_1\cdot \E_{\Pi}(\nabla^2 \kappa_j(x,y)) \cdot\tau^2 + o(\tau^2)
$$
by Lemma~\ref{lem::bias2}.
The variance follows from standard calculation of nonparametric estimation that
$$
V_N(x,y) =\frac{1}{N \tau^2}\cdot C_2\cdot\kappa(x,y)+ o\left(\frac{1}{N\tau^2}\right).
$$
Therefore,
the mean integrated square error is
at rate
$O(\tau^4) + O\left(\frac{1}{N\tau^2}\right)$.
By equating the two rates, we obtain the optimal smoothing
parameter $\tau^* = O(N^{-1/6})$.

\end{proof}

\bibliography{intensity}
\bibliographystyle{amsplain}

\end{document}